\documentclass[aps,pra,reprint,superscriptaddress,longbibliography]{revtex4-1}

\usepackage{amsmath,bm}
\usepackage{graphicx}
\usepackage{dcolumn}
\usepackage{bm}
\usepackage{amssymb}
\usepackage{amsmath}
\usepackage{amsfonts}
\usepackage{epsf}
\usepackage{epsfig}
\usepackage{color}
\usepackage{siunitx}
\usepackage{microtype} 
\usepackage{nicefrac}

\renewcommand{\i}{\textit}
\newcommand{\eref}[1] {(\ref{#1})}
\newcommand{\Eref}[1] {Eq.~(\ref{#1})}
\newcommand{\Fref}[1] {Fig. \ref{#1}}
\newcommand{\Tref}[1] {Table \ref{#1}}

\begin{document}
\title{
Revealing the Two-Electron Cusp in the Ground States of He and H$_2$ \\
via Quasifree Double Photoionization
}

\author{Sven~Grundmann}
\email{grundmann@atom.uni-frankfurt.de}
\affiliation{Institut f\"ur Kernphysik, Goethe-Universit\"at, Max-von-Laue-Strasse 1, D-60438 Frankfurt, Germany \\}
\author{Vladislav~V.~Serov}
\affiliation{Department of Theoretical Physics, Saratov State
  University, Saratov 410012, Russia}
\author{Florian~Trinter}
\affiliation{Photon Science, Deutsches Elektronen-Synchrotron (DESY), Notkestrasse 85, D-22607 Hamburg, Germany}
\affiliation{Molecular Physics, Fritz-Haber-Institut der Max-Planck-Gesellschaft, Faradayweg 4-6, 14195 Berlin, Germany}
\author{Kilian~Fehre}
\author{Nico~Strenger}
\author{Andreas~Pier}
\author{Max~Kircher}
\author{Daniel~Trabert}
\author{Miriam~Weller}
\author{Jonas~Rist}
\author{Leon~Kaiser}
\affiliation{Institut f\"ur Kernphysik, Goethe-Universit\"at, Max-von-Laue-Strasse 1, D-60438 Frankfurt, Germany \\}
\author{Alexander~W.~Bray}
\affiliation{Research School of Physics, Australian National University, Canberra ACT 2601, Australia \\}
\author{Lothar~Ph.~H.~Schmidt}
\affiliation{Institut f\"ur Kernphysik, Goethe-Universit\"at, Max-von-Laue-Strasse 1, D-60438 Frankfurt, Germany \\}
\author{Joshua~B.~Williams}
\affiliation{Department of Physics, University of Nevada, Reno, NV 89557, USA \\}
\author{Till~Jahnke}
\author{Reinhard~D\"orner}
\author{Markus~S.~Sch\"offler}
\email{schoeffler@atom.uni-frankfurt.de}
\affiliation{Institut f\"ur Kernphysik, Goethe-Universit\"at, Max-von-Laue-Strasse 1, D-60438 Frankfurt, Germany \\}
\author{Anatoli~S.~Kheifets}
\email{a.kheifets@anu.edu.au}
\affiliation{Research School of Physics, Australian National University, Canberra ACT 2601, Australia \\}

\date{\today}

\begin{abstract}
We report on kinematically complete measurements and \i{ab initio} non-perturbative calculations of double ionization of He and H$_2$ by a single 800 eV circularly polarized photon.
We confirm the quasifree mechanism of photoionization for H$_2$ and show how it originates from the two-electron cusp in the ground state of a two-electron target.
Our approach establishes a new method for mapping electrons relative to each other and provides valuable insight into photoionization beyond the electric-dipole approximation.
\end{abstract}
\maketitle
\section{Introduction}
Many-electron correlations in atoms and molecules have been a subject of intense theoretical and experimental scrutiny \cite{Yarkony1995}.
One manifestation of such correlations are the so-called cusps, i.e., the points in the coordinate space where the two correlated particles coalesce.
These cusps are fundamental for understanding the photoabsorption process \cite{Suric2003}.
The electron-nucleus cusp is the most prominent one \cite{Kato1957}.
It has a major influence on the total binding energy of the system and is well tested by spectroscopic techniques.
The two-electron cusp is much more subtle.
Only very few highly correlated ground-state wave functions display this cusp correctly \cite{Hylleraas1930,James1933}.
Traditional photoionization studies are not capable of probing it 
because the singular point in the phase space barely contributes to the total cross section.
Indeed, at high (but non-relativistic) energies, the Born approximation demonstrates how the dependence of the cross section on the photon energy $\omega$ characterizes the initial spatial probability density of electrons relative to the nucleus \cite{Bethe1957}.
Accordingly, the total single ionization cross section $\sigma^+$ scales as $Z^{5}/\omega^{7/2}$ for hydrogen-like $^1 S$ orbitals with $Z$ being the nuclear charge.
For two-electron targets, double ionization is facilitated by electron-electron correlation via the shake-off (SO) and two-step-one (TS1) processes \cite{McGuire1997,Samson1990,Knapp2002}.
At high photon energies, the ratio of double-to-single ionization probabilities $\sigma^{2+}/\sigma^{+}$ converges to the so-called shake-off limit, where two-step-one no longer plays a role \cite{Aberg1970,Spielberger1995}.
In this limit, the SO probability becomes a constant fraction of the single ionization cross section.
In SO, double ionization proceeds through the quasi-instantaneous removal of the first electron, whereas the second electron cannot relax adiabatically to the singly charged ionic ground state.
Instead, the secondary electron is either shaken up to a discrete excitation or shaken off to the continuum.
As single ionization is a precursor to SO, this two-electron correlation process also depends on the spatial probability density of electrons relative to the nucleus.
\textcolor{black}{
The double-to-single ionization ratio in the shake-off limit $\sigma^{2+}_\text{SO}/\sigma^{+}$, on the other hand, is determined by the strength of the electron-electron correlation in the initial state.
This correlation can be pictured as the overlap of the electronic clouds that is stronger for He than for H$_2$ because the the major part of the clouds is localized on two spatially separated nuclei in the molecule.
Accordingly, $\sigma^{2+}_\text{SO}/\sigma^{+}$ equals 1.66\% for He \cite{Andersson1993} and 0.7\% for H$_2$ \cite{Siedschlag2005}. 
}

It had been predicted by \citet{Amusia1975} that under certain kinematic conditions, the quasifree mechanism (QFM) facilitates double ionization without any involvement of the nucleus.
QFM leads to the creation of a quasifree electron pair that is emitted back-to-back with equal energy sharing.
Accordingly, the nucleus is only a spectator, remaining nearly at rest because the inter-electron degree of freedom absorbs the energy and momentum of the photon.
Correct weighting of QFM relative to the other one-photon double ionization (PDI) processes requires the two cusp conditions introduced by Kato \cite{Kato1957,Thakkar1976}
\begin{equation*}
{d\rho'(0)/[-2Z\rho(0)]} =1 \ \ \text{and} \ \ {h'(0)/ h(0)}=1
\end{equation*}
to be considered.
Here $\rho(r_{1,2})$ are the single electron densities for electrons 1 and 2 with respect to the nucleus and $\rho'=\mathrm{d} \rho / \mathrm{d} r_{1,2}$.
$h(r_{-})$ is the so-called {\em intracule} \cite{Eddington1946}, i.e., the initial spatial probability density of electrons relative to each other, $r_{-}$ is the inter-electronic distance, and $h'=\mathrm{d} h / \mathrm{d} r_-$.
Note that the shortcut \textit{intracule} is commonly used for the square modulus of the intracule wave function.
Because QFM is most efficient when the two electrons are located close to each other, it can reveal $h(r_-=0)$ and hence the two-electron cusp in the ground state of a two-electron target.
\textcolor{black}{
This relation and an adequate analytical procedure to approximate $h(0)$ through the known QFM cross section are presented in the current work.}

\textcolor{black}{
Recently, the breakdown of the electric-dipole approximation in photoionization has been investigated intensely in the multi-photon and one-photon regimes (e.g., Refs.~\cite{Chelkowski2014,Hartung2019,Grundmann2020b}).
The QFM is a pure electric-quadrupole contribution to one-photon double ionization and thereby a particularly unambiguous example of a nondipole effect.}
The QFM was confirmed experimentally in the helium atom by \citet{Schoffler2013}.
As the ground-state wave functions of He and H$_2$ both have the same $^1 S$ symmetry, the back-to-back emission at equal energy sharing is forbidden by a dipole selection rule \cite{Maulbetsch1995,Weber2004a}.
Accordingly, the QFM can be isolated clearly in a fully differential cross section \cite{Grundmann2018}.
In the present work, we have used this experimental access to confirm the quasifree mechanism for the H$_2$ molecule irradiated with 800~eV circularly polarized photons.

\section{Experimental and numerical techniques}
In our experiments, we employed a COLTRIMS (Cold Target Recoil Ion Momentum Spectroscopy) reaction microscope \cite{Dorner2000, Ullrich2003, Jahnke2004a} and intersected a supersonic jet of the respective target gas with a synchrotron beam of 800~eV photons from beamline P04 at $\text{PETRA III}$ (DESY, Hamburg \cite{Viefhaus2013}).
We used circularly polarized photons because beamline P04 is currently not able to provide linearly polarized light due to a high heat load on the first mirror.
In order to increase the photon flux to an estimated maximum of $1.6 \times10^{14}$ photons/s, we used a so-called pink beam by setting the monochromator to zeroth order.
Additionally, an aluminium blank mirror was used instead of the usual monochromator gratings of beamline P04.
To exclude low-energy photons, a foil filter was inserted into the beam path.
The reaction fragments from the interaction region were guided by electric and magnetic fields towards two time- and position-sensitive detectors \cite{Jagutzki2002,Jagutzki2002a}.
Apart from one electron, we detect all the reaction fragments in coincidence and calculate their three-dimensional momentum vectors from the times-of-flight and positions-of-impact.
The missing electron's momentum vector is calculated using momentum conservation.
This procedure is less accurate for H$_2$, as the center of mass has to be calculated from two protons instead of being directly measured via the doubly charged He$^{2+}$ nucleus.
Thus, the systematic error propagating to the calculated electron is larger and the noise reduction (exploiting energy conservation) is less efficient in case of H$_2$.
The different signal-to-noise ratios explain why the agreement between experiemt and theory is better for He than for H$_2$ in this work.

Absolute cross sections cannot be retrieved from the experimental data and therefore measured differential cross sections for H$_2$ and He cannot be inter-normalized from these datasets alone.
This can be achieved by numerical computations using the external complex scaling method in the prolate spheroidal coordinates (PSECS) \cite{Serov2009}.
Said {\it ab initio} method is based on a solution of the six-dimensional driven Schr\"odinger equation,
\begin{equation}
(\hat{H}_0-E)\Psi^{(+)}(\boldsymbol{r}_1,\boldsymbol{r}_2)= -\hat{H}_{\rm int}
\Phi_0(\boldsymbol{r}_1,\boldsymbol{r}_2) ~,
\end{equation} 
for the first order wave function $\Psi^{(+)}(\boldsymbol{r}_1,\boldsymbol{r}_2)$ with a boundary condition for the outgoing wave.
Here $\boldsymbol{r}_{1,2}$ are the position vectors for electrons 1 and 2 with respect to the nucleus.
$\hat{H}_0$ is an unperturbed two-electron Hamiltonian in the field of the two fixed nuclei and $\Phi_0(\boldsymbol{r}_1,\boldsymbol{r}_2)$ is the initial-state electronic wave function.
Earlier, PSECS has been applied for calculations of dipole PDI \cite{Serov2009,Serov2012}.
Presently, the quadrupole interaction is also included in $\hat{H}_{\rm int}$.
The two-electron Hamiltonian of the non-relativistic electromagnetic interaction in the Poincar\'e gauge, truncated to the quadrupole term, has the form
\begin{equation}
\hat H_{\rm int} =  
\boldsymbol{\epsilon} \cdot (\boldsymbol{r}_1 + \boldsymbol{r}_2) + 
\frac{i}{2} 
\left[(\boldsymbol{\epsilon} \cdot \boldsymbol{r}_1) (\boldsymbol{k}_\gamma \cdot \boldsymbol{r}_1) + (\boldsymbol{\epsilon} \cdot \boldsymbol{r}_2) (\boldsymbol{k}_\gamma \cdot \boldsymbol{r}_2) \right]. \label{Hint}
\end{equation}
Here $\boldsymbol{\epsilon}$ is the polarization vector and $\boldsymbol{k}_\gamma=k_\gamma \boldsymbol{n}_\gamma$ is the photon momentum vector.
Note that the magnetic-dipole and the electric-quadrupole terms are of the same order in the expansion beyond the electric-dipole approximation.
However, the electric-quadrupole term is dominant in the $s$-wave of the relative electron motion which forms the cusp, whereas the magnetic-dipole term contributes mostly to the $p$-wave (see Ref. \cite{wang2020} for further details).
PSECS calculated total integrated cross sections are listed in \Tref{Tab1}.
\textcolor{black}{
The discrepancy between the present result for the quadrupole contribution to the total cross section in He PDI and the one from Ref. \cite{Ludlow2009} is due to the fact that a quadrupole operator proportional to the spherical harmonic $Y_{20}$ was used in the latter work.
When using the same operator with PSECS, the same total integrated cross section value is obtained as in \cite{Ludlow2009}.
However, as shown in Ref. \cite{Grundmann2018}, the quadrupole term in the interaction operator \eref{Hint} should be expressed in terms of $Y_{21}$ to yield the correct photoelectron angular distributions.
The QFM cross section in Ref. \cite{Schoffler2013} was merely estimated to 0.1\% of the total PDI cross section (see Note \cite{Note1} for further details) and we do not know how the QFM cross section was defined here.
In the present work, we define QFM as the peak in the doubly differential cross section around equal energy sharing and back-to-back emission.
Accordingly, the total integrated cross sections of the quasifree mechanism correspond to the shaded areas in \Fref{Fig2} (i.e., a rectangle in \Fref{Fig1} respectively).
}

\begingroup
\setlength{\tabcolsep}{10pt} 
\renewcommand{\arraystretch}{1.5} 
\begin{table*}
\
\centering
\caption{
Total integrated cross sections for single and double ionization of He and H$_2$ by a 800 eV circularly polarized photon.
The present PSECS calculations are compared to various published data.
In the present work, the total integrated cross sections of the quasifree mechanism (QFM) correspond to the shaded areas in \Fref{Fig2}.
\label{Tab1}}
\begin{tabular}{lll|llllll}
 \hline \hline
  (barn)& \multicolumn{2}{c}{Single ionization} & \multicolumn{6}{c}{Double ionization} \\
  & & & \multicolumn{2}{c}{Dipole} & \multicolumn{2}{c}{Quadrupole} & \multicolumn{2}{c}{QFM} \\
  & Present & Ref.~\cite{Yan1998} & Present & Ref.~\cite{Ludlow2009} & Present & Ref.~\cite{Ludlow2009} & Present & Ref.~\cite{Schoffler2013} \\
  He & 730 & 784 & 19.5 & 19.2 & 0.10 & 1.21 & 0.039 & 0.02 \\
  H$_2$ & 62 & 71 & 0.75 & & 0.015 & & 0.008 & \\
  \hline \hline
\end{tabular}
  \label{tab:1}
\end{table*}  
\endgroup

\section{Separating the QFM cross section}
\begin{figure}
\centering
\includegraphics[width=1.0\columnwidth]{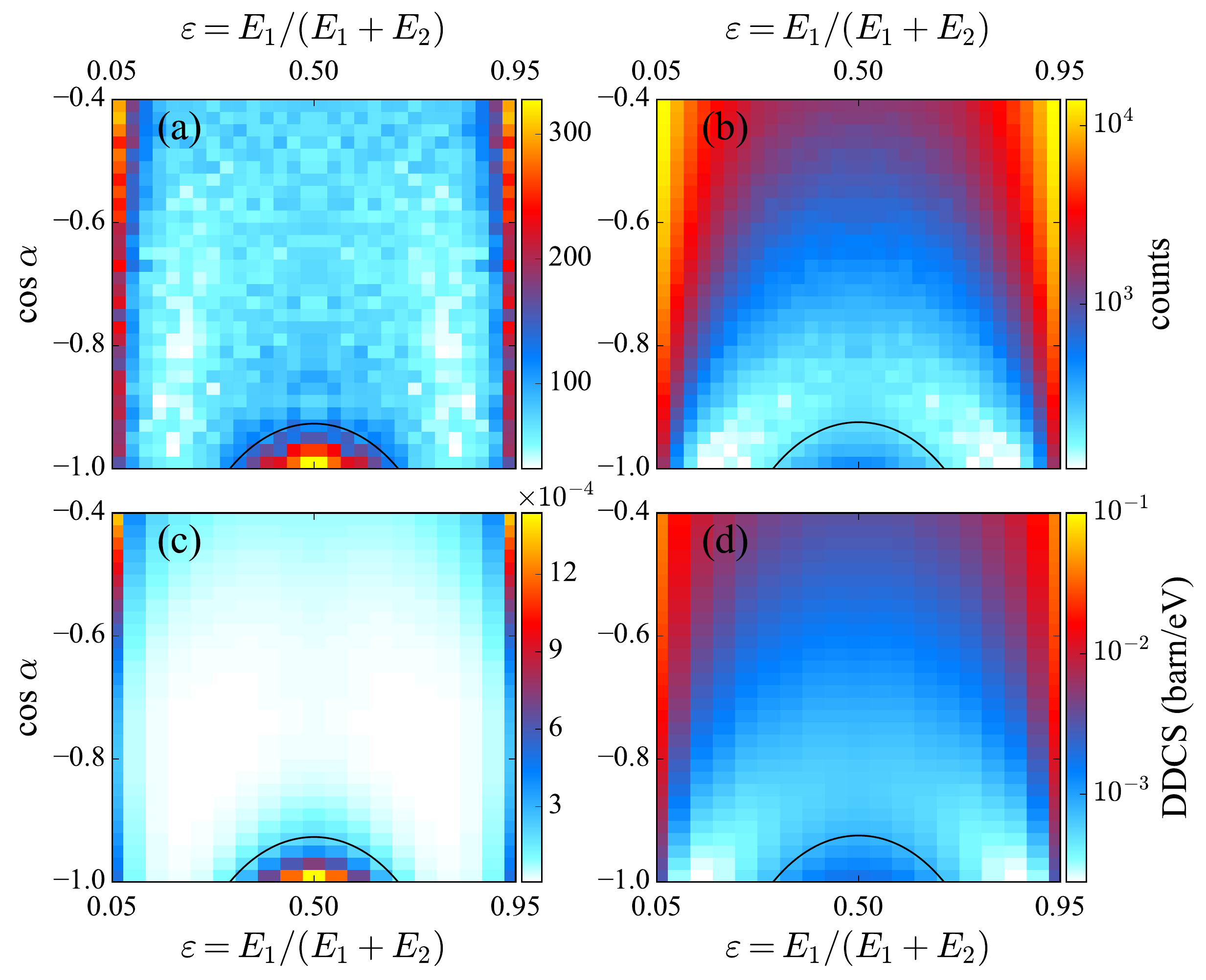}
\caption{
Measured [calculated] doubly differential cross sections [$\mathrm d^2 \sigma (E_1,\alpha)/\mathrm d E_1 \mathrm d \alpha$] of H$_2$ in (a) [(c)] and He in (b) [(d)] for PDI by a single 800 eV circularly polarized photon.
The contributions around equal energy sharing ($\varepsilon =$ 0.5) and back-to-back emission ($\cos\alpha =-1$) correspond to the QFM and are representative of the electron-electron cusp in the two-electron ground state.
In the case of He, the QFM contribution can only be seen against the dipole background with a logarithmic scale display.
The black line indicates the positions in momentum space where $\boldsymbol{K} = 2 \text{ atomic units}$. Note that $\boldsymbol{K} = 0$ at $\varepsilon =$ 0.5 and $\cos\alpha =-1$.
} 
\label{Fig1}
\end{figure}

To search for the QFM fingerprint, the electron mutual angle $\alpha= \cos^{-1} ( \boldsymbol{k}_1 \cdot \boldsymbol{k}_2 / (|\boldsymbol{k}_1| |\boldsymbol{k}_2|))$
is analysed along with the electron energy sharing calculated as $\varepsilon=E_1/(E_1+E_2)$.
Here $\boldsymbol{k}_{1,2}$ and $E_{1,2}$ are the momentum vectors and the kinetic energies of the electrons 1 and 2, respectively.
Figures \ref{Fig1} (a) and (b) show the measured doubly differential cross sections (DDCS) [$\mathrm d^2 \sigma (E_1,\alpha)/\mathrm d E_1 \mathrm d \alpha$] for PDI of H$_2$ and He by a single 800 eV circularly polarized photon.
The events resulting from QFM are located around equal energy sharing ($\varepsilon =0.5$) and back-to-back emission ($\cos\alpha = -1$).
They correspond to (almost) zero recoil momentum of the center of mass.
In comparison to other features, QFM is more intense in H$_2$ than in He, suggesting a higher ratio $\sigma_{\rm QFM}^{2+} / \sigma^{2+}$ in the former target, which is in line with the results presented in Table I.
Figures \ref{Fig1} (c) and (d) show the calculated DDCS for PDI of H$_2$ and He, that are in excellent agreement with the experimental results.
Note that the QFM contribution for He can only be seen against the dipole background in \ref{Fig1} (b) and (d) with a logarithmic scale display.

\begin{figure}
\centering
\includegraphics[width=0.99\columnwidth]{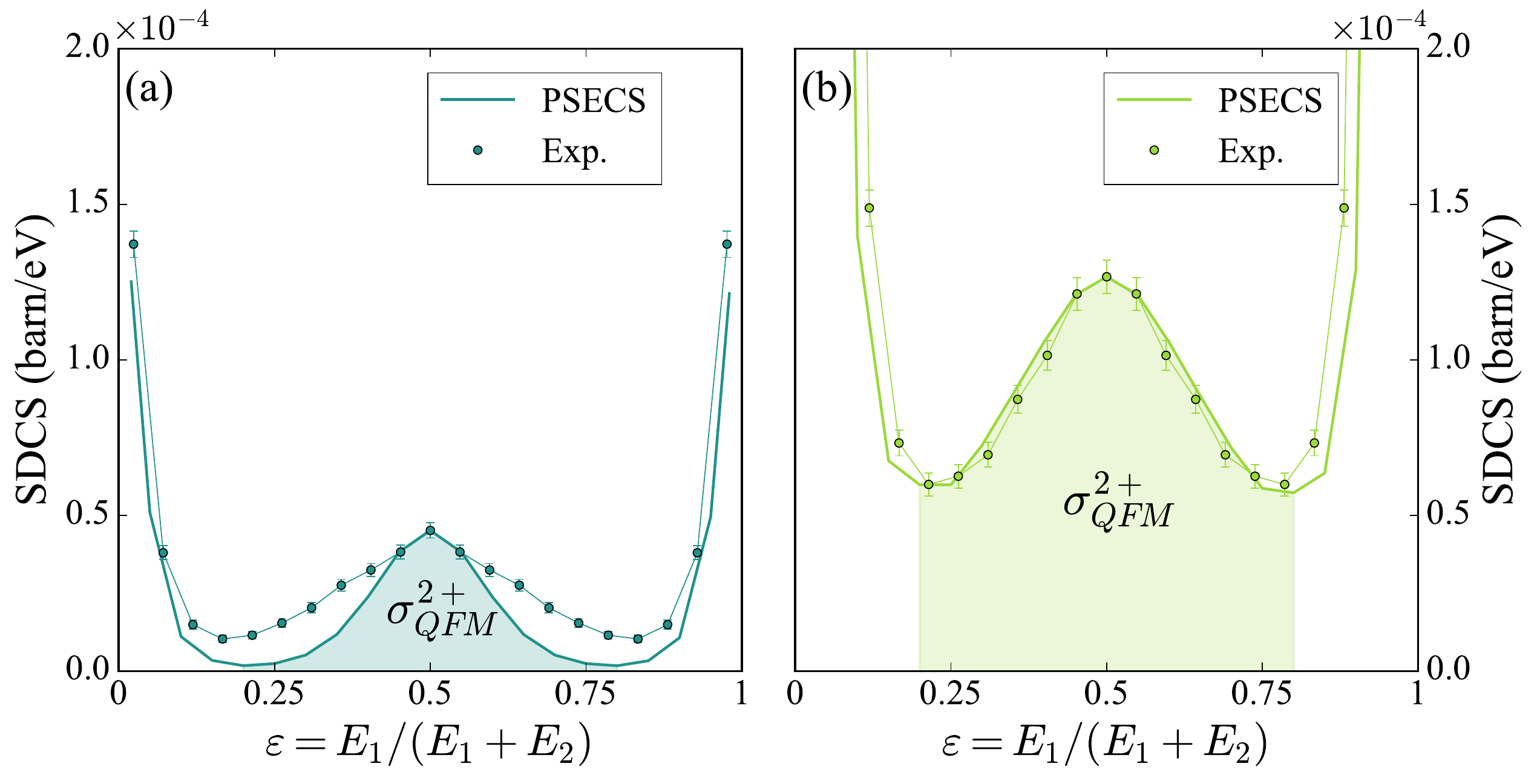}
\caption{
Singly differential cross sections [$\mathrm d\sigma(E_1)/\mathrm d E_1$] for PDI of H$_2$ in (a) and He in (b) by a single 800 eV circularly polarized photon for electrons emitted back-to-back (theory and experimental data are integrated over $\alpha = 180^\circ \pm 30^\circ$).
The experimental datasets are normalized to theory at the equal energy sharing point.  The colored areas under the theory curves represent the QFM cross sections tabulated in \Tref{Tab1}.
} 
\label{Fig2}
\end{figure}

With the kinematically complete experimental data and {\em ab initio} calculations, we can  examine the differences in the correlated structure of the ground states of He and H$_2$.
Figure \ref{Fig2} presents a singly differential cross section (SDCS) for PDI of He and H$_2$, for events from the QFM-dominated range of the electron mutual angle ($\alpha = 180^\circ \pm 30^\circ$) and resolved for the energy of one electron.
The two theory curves share the same absolute scale and the experimental data are normalized to theory at the equal energy sharing point.
The peak distributions around equal energy sharing represent the QFM without any involvement of the nucleus.
As shown in the next section, the strength of the equal energy peak relates to the electron-electron pair density $h(0)$ in the ground-state wave functions of He and H$_2$.
Contrastingly, an asymmetric energy sharing requires a nucleus to compensate the recoil of the two emitted electrons which is imparted by the SO process.
This process dominates the total integrated cross sections of He and H$_2$ PDI at 800 eV photon energy \cite{Knapp2002}.
For SO photoionizaton, a small energy transfer, i.e., a very unequal energy sharing, is strongly favored and the slow electron is emitted almost isotropically \cite{Knapp2002}.
Thus, the probability of SO photoionization depends only weakly on the electron mutual angle $\alpha$.

\begin{figure}
\centering
\includegraphics[width=0.8\columnwidth]{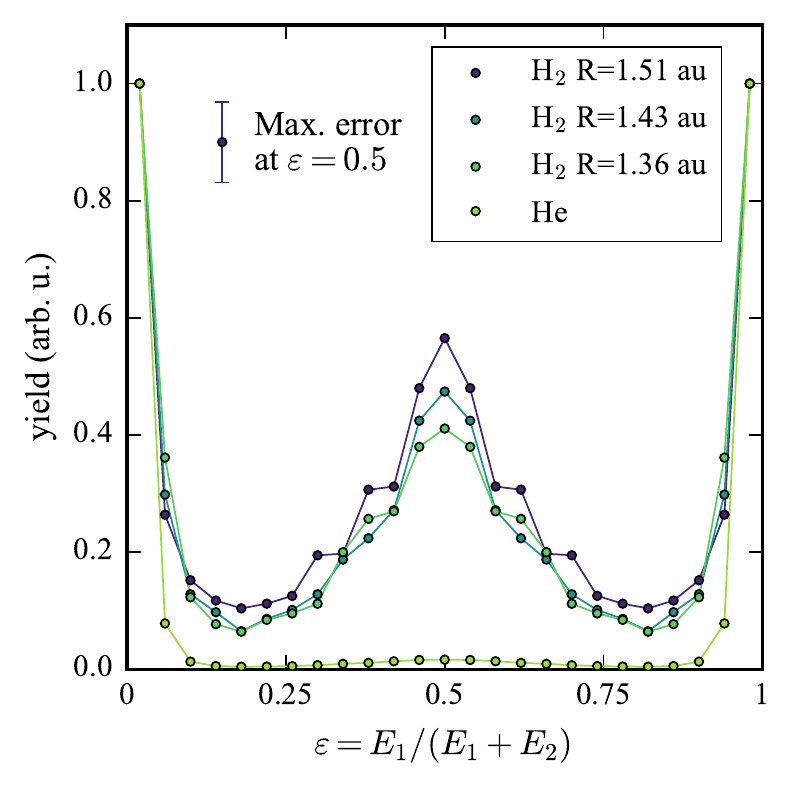}
\caption{
Experimental singly differential intensities [$\mathrm d I(E_1)/\mathrm d E_1$] for PDI of H$_2$ and He by a single 800 eV circularly polarized photon for electrons emitted back-to-back (integrated over $\alpha = 180^\circ \pm 30^\circ$), resolved for different internuclear distances $R$.
The datasets are inter-normalized at the fringes of highly asymmetric energy sharing.
} 
\label{fig_R}
\end{figure}

\textcolor{black}{
Up to now, we considered H$_2$ at the average internuclear distance of $R= 1.4$~au.
Furthermore, we can use the reflection approximation and relate $R$ with the kinetic energy release (KER) via $\text{KER} = 1/ R$ (both quantities are expressed in atomic units).
This way we can investigate differential cross sections depending on $R$ by inspecting subsets of our data for which the KER is in a certain range, as shown in \Fref{fig_R}.
Note that He corresponds to an internuclear distance of $R=0$.
The experimental datasets in \Fref{fig_R} are inter-normalized at the highly asymmetric energy sharing fringes.
By increasing $R$, SO and QFM cross sections decrease in absolute terms as learned from \Fref{Fig2}.
However, SO decreases at a faster rate and the probability of the QFM at the energy sharing midpoint grows relatively to the SO fringes.
Accordingly, Fig. \ref{fig_R} further encourages the following physical interpretation.
}\textcolor{black}{As the internuclear distance $R$ grows, the overlapping potential wells of the two protons as well as the electronic clouds are further separated.
While shallower potential wells lead to a lower $\sigma^{+}$, less electron-electron correlation reduces $\sigma^{2+}_\text{SO}/\sigma^{+}$.
Hence, SO is strongly suppressed via the expansion of the molecule.
For QFM, on the other hand, the decline of the cross section is less pronounced.
A possible intuitive explanation is that the electron-electron cusp is barely affected by a growing $R$ because both electrons stay close to the center point between the two protons to partake in the bonding.
Accordingly, the system accessibility for QFM photoionization remains relatively strong.
}

\section{Connecting the QFM cross section and the intracule}
The relation of the single ionization cross section $\sigma^+$ to the relative position of electrons and nuclei follows from the applicability of the Born approximation.
Analogously, the QFM probability is related to the structure of the intracule wave function as a part of the quadrupole acts directly on the inter-electron relative coordinate (see e.g. \cite{Ludlow2009, Dorner1996}).
To demonstrate this relation, we introduce the Jacobian coordinates and their conjugate momenta:
\begin{eqnarray*}
\boldsymbol{r}_-=\boldsymbol{r}_1-\boldsymbol{r}_2,~
&& 
\boldsymbol{r}_+=(\boldsymbol{r}_1+\boldsymbol{r}_2)/2,
\\
\boldsymbol{k}=(\boldsymbol{k}_1-\boldsymbol{k}_2)/2, \text{ and }
&&
\boldsymbol{K}=(\boldsymbol{k}_1+\boldsymbol{k}_2) 
\ .
\end{eqnarray*}
Here $\boldsymbol{r}_-$ and $\boldsymbol{k}$ describe the relative electron motion whereas $\boldsymbol{r}_+$ and $\boldsymbol{K}$ are related to the electron-pair center of mass.
In these variables, the interaction operator \eref{Hint} takes the form
\begin{equation}
\hat H_{\mathrm{int}} = 2\boldsymbol{\epsilon} \cdot \boldsymbol{r}_+ + i
(\boldsymbol{\epsilon} \cdot \boldsymbol{r}_+) (\boldsymbol{k}_\gamma \cdot \boldsymbol{r}_+) + \frac{i}{4} 
(\boldsymbol{\epsilon} \cdot \boldsymbol{r}_-) (\boldsymbol{k}_\gamma \cdot \boldsymbol{r}_-) \ .
\label{eq1}
\end{equation}
The first term is the electric dipole (E1) contribution to the transition amplitude, the second and third term contain the electric quadrupole (E2) contribution.
While the dipole acts only on the ``$+$'' coordinate, transferring the recoil to the center of mass, the part of the quadrupole
\begin{eqnarray}
\hat H_{-} &=& \frac{ik_\gamma}{4} (\boldsymbol{\epsilon} \cdot \boldsymbol{r}_-)
(\boldsymbol{n}_\gamma \cdot \boldsymbol{r}_-) \ 
\label{H_QFM}
\end{eqnarray}
acts directly on the inter-electron separation (the ``$-$'' coordinate).
When the electrons are emitted back-to-back with equal energy, they balance each other's momentum.
Accordingly, as nuclear recoil is not involved, this part of the quadrupole contribution is responsible for the QFM.\\

\begin{figure}
\includegraphics[width=0.8\columnwidth]{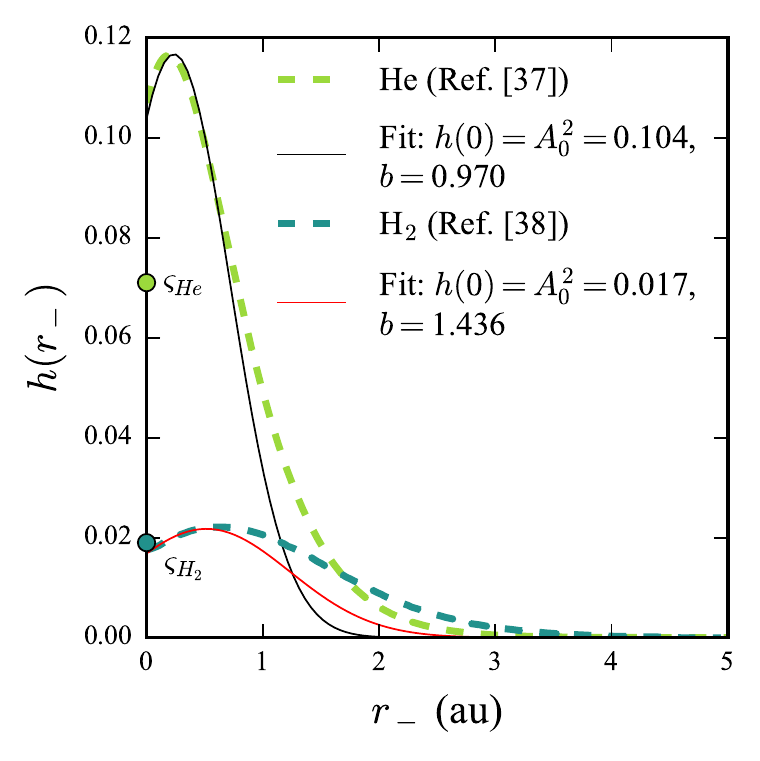}
\caption{
Intracules of He \cite{Thakkar1977} and H$_2$ \cite{Koga1993} fitted with ansatz \eref{intra}.
The circles present approximations of $h(0)$ for the two targets obtained from applying \Eref{varsigma} on the PSECS cross sections.
}
\label{intracule}
\end{figure}

For a more qualitative analysis, we consider the ground-state wave function of the two electrons in the following form
\begin{eqnarray} \nonumber
\Phi_0(\boldsymbol{r}_+,\boldsymbol{r}_-)
&=& \chi_0(\boldsymbol{r}_+)\psi_0(r_-) \ .
\label{Phi0}
\end{eqnarray}
Here the ground-state wave function of relative motion (the intracule wave function) is
\begin{eqnarray}
\psi_0(r_-) &=& \frac{1}{\sqrt{4\pi}} A_0 \exp [ r_-/2 - r_-^2/b^2 ]~,
\label{intra}
\end{eqnarray}
and $\chi_0(\boldsymbol{r}_+)$ is the extracule wave function \cite{Eddington1946}.
%
The intracule wave function Eq. \eref{intra} is chosen to satisfy the cusp condition at $r_-\to 0$.
The Gaussian multiplier with the cut-off parameter $b$ is introduced to compensate an infinite growth of the exponential multiplier as $r_-\to\infty$. 
As shown in \Fref{intracule}, the intracule $h(r_-)=4\pi|\psi_0(r_-)|^2$ has the form of a shifted Gaussian which approximates the intracules of He \cite{Thakkar1977} and H$_2$ \cite{Koga1993} quite accurately.
Accordingly, the amplitude of the QFM process can be written in the form
\begin{eqnarray}
f_{\rm QFM} &=&  (2\pi)^{-3}
\langle e^{i \boldsymbol{k}_1 \cdot \boldsymbol{r}_1+ i \boldsymbol{k}_2 \cdot \boldsymbol{r}_2}|\hat{H}_-|\Phi_0\rangle \nonumber \\
 &=& \frac{ik_\gamma}{4} f_+(\boldsymbol{K}) f_-(\boldsymbol{k})~,
\end{eqnarray}
where 
\begin{eqnarray}
\label{amplitudes_1} 
f_+(\boldsymbol{K}) &=& (2\pi)^{-3/2}
\langle e^{i \boldsymbol{K} \cdot \boldsymbol{r}_+} | \chi_0(\boldsymbol{r}_+)\rangle ~,
\\
\label{amplitudes_2} 
f_-(\boldsymbol{k}) &=& (2\pi)^{-3/2} \langle e^{i \boldsymbol{k} \cdot \boldsymbol{r}_-}|(\boldsymbol{\epsilon} \cdot \boldsymbol{r}_-)
(\boldsymbol{n}_\gamma \cdot \boldsymbol{r}_-)|\psi_0(r_-)\rangle \nonumber\\
&=& \frac{12\sqrt{2}}{\pi} \frac{(\bm{\epsilon}\cdot\boldsymbol{k}) (\boldsymbol{n}_\gamma\cdot\boldsymbol{k})}{k^2} \frac{A_0}{k^6}
+ O(k^{-8})~.
\end{eqnarray}

The normalization constant $|A_0|^2=h(0)$ is expressed via the intracule at $r_-=0$ alone.
By using Eqs. (\ref{amplitudes_1}) and (\ref{amplitudes_2}), the differential QFM cross section acquires the asymptotic form
\begin{eqnarray}
&& \sigma^{2+}_{\rm QFM}(\boldsymbol{k}_1,\boldsymbol{k}_2)= \frac{4\pi^2\omega}{c} k_1 k_2 |f_{\rm QFM}|^2  \nonumber \\
&& = \frac{72\omega^3}{c^3E^5} \rho(\boldsymbol{k},\boldsymbol{K})h(0)g(\boldsymbol{K}) + O(\omega^{-3})~,
\label{QFM_h0gK} 
\end{eqnarray}
where $c$ is the speed of light and $E = E_1+E_2$. 
Here, we introduced 
\begin{equation*}
g(\boldsymbol{K})=(2\pi)^{-3}|\langle e^{i \boldsymbol{K} \cdot \boldsymbol{r}_+}|\chi_0(\boldsymbol{r}_+)\rangle|^2~,
\end{equation*}
which is the momentum distribution of the electron-pair center of mass in the ground state, and the dimensionless function
\begin{equation*}
\rho(\boldsymbol{k},\boldsymbol{K}) = \frac{|(\bm{\epsilon}\cdot\boldsymbol{k}) (\boldsymbol{n}_\gamma\cdot\boldsymbol{k})|^2}{k^4} \frac{E^5|\boldsymbol{k}+\boldsymbol{K}/2| |\boldsymbol{k}-\boldsymbol{K}/2|}{k^{12}}~.
\label{rho_def}
\end{equation*}

\textcolor{black}{
Equation (\ref{QFM_h0gK}) connects the two-electron pair density to the QFM cross section.}
However, the dependence on the momentum extracule $g(\boldsymbol{K})$ makes this relation less straightforward.
Hence, to retrieve the two-electron cusp $h(0)$ from the QFM cross section, we introduce the proportional-to-intracule cross section
integral (PICSI) which does not depend on the extracule.
For this purpose, we use the normalization condition
\begin{equation*}
\oint \int g(\boldsymbol{K}) K^2 dK d\Omega_K = 1 ~,
\label{Norm1}
\end{equation*}
and once we integrate the value 
\begin{equation*}
\sigma^{2+}_{\rm QFM}(\boldsymbol{k}_1,\boldsymbol{k}_2)/\rho(\boldsymbol{k},\boldsymbol{K})\propto h(0)g(\boldsymbol{K})
\end{equation*}
over $\boldsymbol{K}$ in the region $K<K_{\rm QFM}$ (where QFM dominates in \Fref{Fig1}) we should get the desired PICSI.

\textcolor{black}{
For He, the momentum extracule is spherically symmetric, i.e., $g(\boldsymbol{K})=g(K)$.
For non-oriented H$_2$, on the other hand, the cross section is proportional to the momentum extracule averaged over all orientations of the molecular axis,
\begin{equation*}
g(K) = \frac{1}{4\pi} \oint g(\boldsymbol{K}) d\Omega_{\boldsymbol{R}},
\label{Norm1}
\end{equation*}
which is also spherically symmetric.\\
In order to attain the doubly differential PDI cross sections as presented in Fig. 1, we integrate $\sigma^{2+}_{\rm QFM}(\boldsymbol{k}_1,\boldsymbol{k}_2)$ over all angles, except for the electron mutual angle $\alpha$, and get
\begin{eqnarray*}
\sigma^{2+}_{\rm QFM}(E_1,\alpha) &=& \oint \int_{0}^{2\pi} \sigma^{2+}_{\rm QFM}(\boldsymbol{k}_1,\boldsymbol{k}_2) d\Omega_1 d\phi_{12} \\ 
&=& \frac{8\pi^2}{15} \frac{72\omega^{3}}{c^3E^5}  \rho(\varkappa,\beta) h(0)g(K)~,
\end{eqnarray*}
where $\phi_{12}$ is the azimuthal angle of $\boldsymbol{k}_2$ projected on the plane perpendicular to $\boldsymbol{k}_1$.
Here, we introduced  
\begin{equation*}
\rho(\varkappa,\beta) = {\sqrt{1-\beta^2}}/{(1-\varkappa)^{6}}~
\end{equation*}
which resembles $\rho(\boldsymbol{k},\boldsymbol{K})$ averaged over all angles except $\alpha$,
where $\varkappa=K^2/4E$ and $\beta=(E_1-E_2)/E=2\varepsilon-1$.\\
To express the PICSI in terms of $\sigma^{2+}(E_1,\alpha)$, we go from single integration over $K$ to double integration over the electron energy sharing $\epsilon$ and the electron mutual angle $\alpha$.
We make use of the identity
\begin{eqnarray*}
&& \int_0^{K_{QFM}} g(K) K^2 dK \equiv \\
&& \int_{0}^{\beta_{QFM}} \int_{-1}^{\eta_{\rm QFM}} g\left(K(\beta,\eta)\right) w(\varkappa) J(\beta,\eta) d\eta d\beta
\end{eqnarray*}
where $\eta=\cos\alpha$, the weight factor is} 
\begin{equation*}
w(\varkappa)={\frac{E}{2}}\max\left(1,\frac{2\varkappa}{\eta_{\mathrm{QFM}}+1}\right) 
\end{equation*}
and the Jacobian reads
\begin{equation*}
J(\beta,\eta)=({E^{1/2}|\eta\beta|})/({4\varkappa^{1/2}\sqrt{1-\beta^2}})~.
\end{equation*}
\textcolor{black}{
Here, $\eta_{\rm QFM}$ and $\beta_{\rm QFM}$ substitute $K_{\rm QFM}$ in confining the QFM-dominated area of the cross section.
The final form of the PICSI is
\begin{align}
\label{varsigma}
\varsigma =& \frac{5c^3E^{5}}{48\pi\omega^{3}} \times \nonumber \\
& \int_{0}^{\beta_{\rm QFM}} \int_{-1}^{\eta_{\rm QFM}}
 \frac{\sigma^{2+}(E_1,\alpha)w(\varkappa)J(\beta,\eta)}{\rho(\varkappa,\beta)} \mathrm{d}\eta \mathrm{d}\beta 
~.
\end{align}
}
Once integrated, \Eref{varsigma} yields $\varsigma_{\rm He}=0.071$ and $\varsigma_{{\rm H}_2}=0.019$ (see \Fref{intracule}).
\textcolor{black}{
Due to the approximations used in the analytical derivation of \Eref{varsigma}, the good agreement for H$_2$ is surprising, and the results for He are a better estimate for the accuracy of the extraction protocol.
}
However, using measured or calculated fully differential double ionization cross sections and following this simple analytical approach, the PICSI yields a good approximation for $h(0)$ of a two-electron target.\\

\section{Conclusion}
\textcolor{black}{
We have confirmed the quasifree mechanism of one-photon double ionization for H$_2$ at 800 eV photon energy.
By comparing differential cross sections for H$_2$ and He PDI, the QFM allows studying the fine details of electron correlation in the ground states of these two targets in the high-photon-energy regime.
Similarly to single photoionization, which reveals the one-electron charge density, the QFM relates to the electron pair density or the squared intracule wave function.
%
This is important because accurate charge densities and intracules are needed for evaluation of x-ray scattering form-factors and intensities.
The latter can be computed from the Fourier transforms of $h(r_-)$ \cite{Benesch1970,Thakkar1976}.
Finally, nearly 50 years since the theoretical prediction of QFM \cite{Amusia1975}, not only has it been confirmed experimentally \cite{Schoffler2013,Grundmann2018}, but has also become a novel tool for many-electron spectroscopy of correlated states of matter.\\
}

\begin{acknowledgments}
S.~G. acknowledges travel support by the Wilhelm and Else Heraeus Foundation and wishes to thank the Australian National University for hospitality.
We acknowledge DESY (Hamburg, Germany), a member of the Helmholtz Association HGF, for the provision of experimental facilities.
Parts of this research were carried out at PETRA III and we would like to thank J\"orn Seltmann and Kai Bagschik for excellent support during the beam time.
We acknowledge support by DFG and BMBF.
\end{acknowledgments}
%

\end{document}